\documentclass[aps,prb,twocolumn,showpacs,groupedaddress]{revtex4}
\usepackage{amsbsy,amssymb,amsmath,bm}
\usepackage{graphicx}

\input{epsf}
\begin{document}

\title{Macroscopic Coulomb blockade in large Josephson junction arrays}
\author{M. V. Fistul$^1$, V. M. Vinokur$^2$, and T. I.  Baturina$^{3}$}
 \affiliation{$^1$Theoretische Physik III,
Ruhr-Universit\"at Bochum, D-44801 Bochum, Germany\\
$^2$Material Science Division, Argonne National
Laboratory, Argonne, Ill. 60439 USA\\
$^3$Institute of Semiconductor Physics, 630090, Novosibirsk, Russia}
\date{\today}

\begin{abstract}

We investigate theoretically transport properties of one- and
two-dimensional regular Josephson junction arrays (JJAs) in an
insulating state. We derive the low-temperature current-voltage
characteristics (the $I$-$V$ dependencies) for the current mediated
by the Cooper pair transfer across the system. In the case where the
screening length $\lambda_c$ associated with the capacitance of the
islands to the ground is much larger than the island's size $d$, we
find that transport is governed by the \textit{macroscopic Coulomb
blockade effect} with the gap $\Delta_c$ well exceeding a single
island charging energy $E_c$. In the limit of $\lambda_c\gg L$,
where $L$ is the linear size of the array, the gap establishes the
dependence on the array size, namely, $\Delta_c\simeq E_c(L/d)$ in
1D and $\Delta_c\simeq E_c\ln(L/d)$ in 2D arrays. We find two
transport regimes: at moderate temperatures, $E_c<k_BT<\Delta_c$,
the low bias transport is thermally activated with the resistance
$R\propto\exp(T_0/T)$ where the activation energy
$k_{\scriptscriptstyle B}T_0=\Delta_c$. At ultra-low temperatures,
$k_{\scriptscriptstyle B}T<E_c$, a JJA falls into a superinsulating
state with $R\propto\exp[(\Delta_c/E_c)\exp(E_c/T)]$.

\end{abstract}

\maketitle

\section{Introduction}

A theoretical and experimental study of large regular Josephson
junction arrays (JJAs), the systems comprised of small
superconducting islands connected by Josephson junctions, has a long
history~\cite{Mooij1,Mooij2,Mooij3,Mooij4,Haviland1,Haviland2,Tighe,Japan1,
Japan2,Fazio,WeesDual,Valles1,Valles2},
see~\cite{Review1,2DJJA_Review} for a review. The remarkable feature
of these systems is that they experience a
superconductor-to-insulator transition (SIT) as the Coulomb charging
energy of a single island, $E_c$ (i.e.  an energy cost to place a
Cooper pair on such an island) compares to the Josephson coupling
energy, $E_{\scriptscriptstyle J}$ measuring the strength of the
phase coupling between the superconducting islands comprising a JJA
~\cite{Anderson64,Abeles77}. In arrays near the critical balance
between $E_c$ and $E_{\scriptscriptstyle J}$, a SIT can be induced
by the magnetic field\cite{noteMagn}. Studies on the
superconductor-insulator transition in thin granular films, which
are often modeled as JJAs, revealed the similar
behavior~\cite{Dynes78,Efetov,White,Sugahara,Orr86,SugaharaLT18,Sugahara26L,
Widom,Jaeger,Barber90,Wu94,GoldmanGranular,Frydman,Barber06}. Even
more remarkably, critically disordered homogeneous superconducting
films exhibited all the wealth of phenomena related to
superconductor-insulator transition characteristics to granular
superconducting
systems~\cite{HavGold,Hebard,ShaharOvadyahu,LiuGoldman,KowalOvadyahu,
VFGIns,Be1,BeInsPMR,BeBSIT,BeDSIT,Shahar-act,Shahar-Coll,TBJETPL,TBPRLQM,Hadacek,
BaturinaIns,BaturinaPhysC,Ovadyahu}.
This brought about the idea that strong disorder induces an
inhomogeneous spatial structure of isolated superconducting islands
in thin homogeneously disordered
films~\cite{LarkinOvchin71n1,IoffeLarkin,MaLee,Imry,KowalOvadyahu,VFGIns}.
Numerical simulations of the homogeneously disordered
superconducting films confirmed that indeed in the high-disorder
regime, the system breaks up into superconducting islands separated
by an insulating sea~\cite{GhosalPRL,GhosalPRB,DubiNat}.  Recent
scanning tunneling microscope measurements of the local density of
states in thin TiN films~\cite{BaturinaSTM08} offered strong support
to this hypothesis.

All the above together shows that the Josephson junction arrays
offer a useful generic model that captures most essential features
of the superconductor-insulator transition in a wide class of
systems ranging from artificially manufactured Josephson junction
arrays to superconducting granular systems and even the
homogeneously disordered superconducting films (see also
review~\cite{Dagotto}) and allows for consideration of all of them
on the common ground. Indeed, recent theoretical results describing
insulating behavior of regular JJAs, appeared to be in a striking
quantitative accordance with the experimental findings in TiN and
InO superconducting disordered films~\cite{FVB,VinNature}.

The common ``working tool" for experimental study of the SIT is the
measurements of transport characteristics of the systems in question.
Altering various parameters of the system, such as tunnel resistance
and transmittance in JJAs,  conditions of deposition, chemical
composition, and thickness of the films, and applied magnetic field,
one can drive the system directly from the superconducting to
insulating state. The transition is observed as a set of the
fan-shaped temperature dependences of the resistance $R(T)$, see
Refs.~[\onlinecite{Mooij1,Mooij3,Mooij4,Haviland2,Japan2,Valles1,Valles2,Dynes78,White,
SugaharaLT18,Jaeger,Barber90,Wu94,GoldmanGranular,Frydman,Barber06,HavGold,Hebard,
ShaharOvadyahu,LiuGoldman,VFGIns,Be1,BeInsPMR,BeBSIT,BeDSIT,
Shahar-act,Shahar-Coll,
TBJETPL,TBPRLQM,Hadacek,BaturinaIns,BaturinaPhysC,Ovadyahu}], with
the activation behavior
\begin{equation}
R(T)~\propto~e^{T_0/T} \label{activDep}
\end{equation}
on the insulating side of the transition, see
Refs.~[\onlinecite{Haviland1,
Tighe,Japan1,Valles1,Valles2,Dynes78,ShaharOvadyahu,LiuGoldman,
KowalOvadyahu,VFGIns,Shahar-act,Hadacek,BaturinaIns,BaturinaPhysC,Ovadyahu}],
where $T_0$ is the activation temperature. Most pronounced features
of this transition manifest themselves in the current-voltage
characteristics (the $I$-$V$ curves). On the superconducting side a
system has very low resistance at low currents followed by jump in
resistivity when current exceeds the critical value. On the
insulating side the $I$-$V$ characteristics show a mirror behavior:
extremely high resistance at low voltages and abrupt jump in the
conductivity at the threshold voltage $V_{\scriptscriptstyle T}$,
see
Refs.~[\onlinecite{Mooij1,Mooij3,Mooij4,Haviland1,Haviland2,Tighe,SugaharaLT18,Sugahara26L,
Jaeger,Barber90,Wu94,GoldmanGranular,Barber06,BeInsPMR,Shahar-Coll,BaturinaIns,
BaturinaPhysC}]. Yet the most startling observations that come from
the insulating side of the transition are: (i) the size dependent
activation energy~\cite{Ovadyahu}; (ii) \textit{hyperactivation}
temperature behavior of resistivity at ultra-low temperatures in
JJAs~\cite{Tighe,Japan1}; (iii) size dependent threshold
voltage~\cite{Mooij1,Haviland2,Baturinatbp}; (iv) peculiar
interrelated magnetic field dependencies of the activation energy
 and threshold voltage in JJAs~\cite{Haviland1} and
superconducting films~\cite{BaturinaIns,BaturinaPhysC}.

The above striking findings called for a theory capable of
quantitative description of the accumulated wealth of the
experimental results within a unified picture. The preceding
publications~\cite{FVB,VinNature} offered such a description in a
framework of the Cooper pair transport in large JJAs in the
insulating region, where $E_c\gg E_{\scriptscriptstyle J}$,
$E_{\scriptscriptstyle J}=\hbar I_c/2e$ ($I_c$ is the critical
current of a single Josephson junction). This paper is the extended
version of earlier publications~\cite{FVB,VinNature}, presenting
the details of the derivation of the current-voltage
characteristics.

We briefly summarize our results: The insulating behavior of the
large Josephson junction array is governed by the
\textit{macroscopic Coulomb blockade effect} with the Coulomb
blockade activation energy
   \begin{align}
        \label{Deltac}\Delta_c = \left\{
          \begin{array}
          [c]{lr}%
          E_c[\Lambda/(2d)]\, \,  ,& 1D \, \text{array} \\
          E_c\ln(\Lambda/d)\, \, ,& 2D \, \text{array}
          \end{array}
          \right.  ,
   \end{align}
where $\Lambda=min\{L,\lambda_c\}$, $L$ is the size of an array, $d$
is the size of the elemental unit of JJA, and $\lambda_c$ is the
screening length related to the capacitance to the ground.
Importantly, this activation energy can be much larger than the
single junction charging energy $E_c$. In the two-dimensional array
the charge binding-unbinding Berezinskii-Kosterlitz-Thouless like
transition takes place at $T=T_{\scriptscriptstyle SI}\simeq
E_c/k_B$ separating the insulating phase existing in the interval
$E_c<k_BT<\Delta_c$ and exhibiting the thermally activated
resistivity (1) with $k_BT_0\equiv\Delta_c$, and the
\textit{superinsulating} state at $T<E_c/k_B$, with
$R\propto\exp[(\Delta_c/E_c)\exp(E_c/k_BT)]$.  In the one
dimensional arrays we expect a crossover between these two states.

The paper is organized as follows: in Section II we introduce the
model and present the general equation for the dc $I$-$V$
dependence, expressing it through the time-dependent correlation
function $K(t)$ of superconducting order parameter phases of the
whole system.  In Sections III and IV we discuss the properties of
$K(t)$ and the corresponding $I$-$V$ characteristics above and below
$T_{\scriptscriptstyle SI}$ respectively.
Section V presents the discussion of the
obtained results.

\bigskip
\section{Model and CVC calculation}

\begin{figure}[tbp]
\includegraphics[width=2.2in]{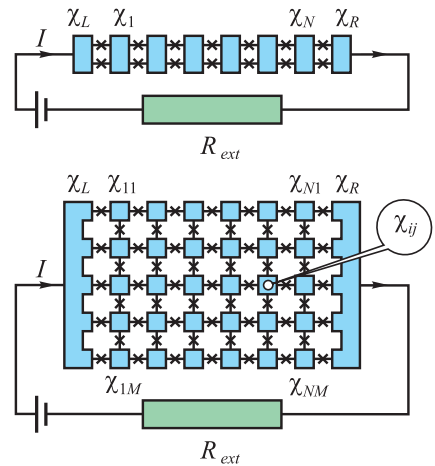}
\caption{Sketch of the considered array geometries. An external
current $I$ is injected from the left through the electrode having
the superconducting phase $\chi_{\scriptscriptstyle L}$ and
extracted through the right electrode with the phase
$\chi_{\scriptscriptstyle R}$.  Upper panel: One-dimensional array
of $N$ superconducting islands (squares) connected by two Josephson
junctions (crosses) to neighbors corresponding to experimental
system of Ref.\,[\onlinecite{Haviland1}]. Lower panel: Two-dimensional $N\times M$
Josephson junction array. } 
\label{circuit}
\end{figure}
The current-voltage characteristics of Josephson systems in an
insulating state have been a subject of intense discussions over the
decades. Single junctions were considered
in~\cite{LikhAverin,Ingold,Shoenrev,KovFistUst,Kuzmin,Kuzmin2} and
the behavior of two-junction systems was studied
in~\cite{Matveev,Zorin,AverinLikharev}. Only a few works were
addressing the Cooper-pair current in large
JJAs~\cite{Efetov,Sugahara,Bradley}.

In this paper we will discuss Cooper pair transport (Josephson
current) and the corresponding $I$-$V$ characteristics in large
JJAs, leaving the calculation of the quasiparticle contribution to
the forthcoming publication. This implies, in particular, that we
neglect interactions of the internal phases with the thermal bath,
since the latter is equivalent to switching on quasiparticle
current.  Let us consider $N\times M$ superconducting islands
comprising a one- ($M=1$) or two-dimensional array closed by a small
(as compared to the quantum resistance for Cooper pairs
$R_{\scriptscriptstyle CP}=h/4e^2\simeq 6.45$\,k$\Omega$) external
resistance, $R_{ext}$, see Fig.\,1. Note, that in this kind of
circuits with a small external load resistance both, a single
Josephson junction~\cite{Ingold,Shoenrev,KovFistUst} and two
Josephson junctions in series~\cite{Zorin} are always in a
superconducting state irrespectively to relation between $E_c$ and
$E_{\scriptscriptstyle J}$. We are interested in a low-temperature
transport, $T\ll T_c$, where $T_c$ is the critical temperature of a
single superconducting island, and, therefore, we can neglect the
fluctuations of the amplitude of the order parameter. We assign the
fluctuating order parameter phase $\chi_{ij}(t)$ to the $\{i,j\}$-th
superconducting island (see Fig.\,1). Josephson relation connects
these phases with the fluctuating voltage drops between the adjacent
islands. We denote the phases of the left- and right leads as
$\chi_{\scriptscriptstyle L}(t)$ and $\chi_{\scriptscriptstyle
R}(t)$, correspondingly. The finite voltage $V$ applied to a JJA
generates the alternating Josephson currents proportional to
$E_{\scriptscriptstyle J}\sin(Vt+\{{\chi}_{ij}\}(t))$. The dc
component of the Josephson current results from the time averaging
of the ac currents and is thus determined by the correlations of the
time-dependent fluctuations of the Josephson phases
$\{{\chi}_{ij}\}(t)$ across the array.

We discriminate between the Josephson coupling energies of intrinsic
junctions $\tilde{E}_{\scriptscriptstyle J}$ and the Josephson
coupling energies $E_{\scriptscriptstyle J}$ between leads and first
and $N$-th rows. Singling out the terms containing the leads phases
$\chi_{\scriptscriptstyle L}$ and $\chi_{\scriptscriptstyle R}$
explicitly we present the array Hamiltonian as:
\begin{widetext}
\begin{eqnarray}\nonumber
H=H_0+H_{bath}+H_{int}\{\chi_{\scriptscriptstyle
L}-\chi_{\scriptscriptstyle R}\} \nonumber
&+&\frac{\hbar^2}{4E_c}\sum_{j=1}^M[(\dot{\chi}_{1j}(t)-\dot
\chi_{\scriptscriptstyle L})^2 +(\dot{\chi}_{Nj}(t)-\dot \chi_{\scriptscriptstyle R})^2 ] 
\\&-&E_{\scriptscriptstyle J}\sum_{j=1}^M
\bigg \{\cos\left[\chi_{\scriptscriptstyle L}(t)-
\chi_{1j}(t)\right]+\cos\left[\chi_{\scriptscriptstyle R}(t)-
\chi_{Nj}(t)\right] \bigg \}\,. \label{Hamilton-predv}
\end{eqnarray}
Here
\begin{equation}
H_0=\sum_{\langle
ij,kl\rangle}\Bigl[\frac{\hbar^2}{4E_c}(\dot{\chi}_{ij}-\dot{\chi}_{kl}-2eV_{ij-kl}/\hbar)^2-\tilde{E}_{\scriptscriptstyle
J}\cos(\chi_{ij}-\chi_{kl})\Bigr]
+\sum_{ij}\frac{\hbar^2}{4E_{c0}}\dot{\chi}_{ij}^2\,, \label{ham}
\end{equation}
\end{widetext}
the brackets $\langle ij,kl\rangle$ denote summation over the pairs
of adjacent junctions, and the last term in (\ref{ham}) represents
the self-charge energies of superconducting islands. We introduced
here the dc voltage drops on the junctions $V_{ij-kl}$. The charging
energies $E_c$ and $E_{c0}$ are determined by the junction
capacitance $C$ and capacitance to the ground $C_0$, as $E_c=2e^2/C$
and $E_{c0}=2e^2/C_0$, respectively. The $H_{bath}$ is the
Hamiltonian characterizing the thermal bath, which can be modeled as
a set of harmonic oscillators with coordinates
$\xi_i$~\cite{Ingold2,Leggett}, i.e.
\begin{equation}
H_{bath}=\sum_i \frac{\hbar^2}{2M_i}[\dot{\xi}_i]^2+M_i\omega_i^2
\xi_i^2\,, \label{Bath}
\end{equation}
where $\omega_i$ and $M_i$ are the frequency and mass of harmonic
oscillators. The $H_{int}$ term in (\ref{Hamilton-predv}) describes
a bilinear coupling of \emph{phases on the leads} to the thermal
heat bath as
\begin{equation}
H_{int}=\sum_i A_i\xi_i (\chi_{\scriptscriptstyle
L}-\chi_{\scriptscriptstyle R})~~, \label{InteractionBath}
\end{equation}
where $A_i$ are the coupling constants. In the circuits
presented in Fig.\,1, the coupling constants $A_i$ are determined
by the external resistance \cite{Shoenrev,Ingold2} $\sum_i
A_i^2/(M_i\omega_i^2)~\simeq~R_{\scriptscriptstyle CP}/R_{ext}$.
Since we consider Josephson currents only, we will neglect the
phase coupling to the thermal heat bath in the internal part of the
array: dissipation on the internal islands implies the presence of
the current of quasiparticles. This means that we can treat the
evolution of the internal phases as a non-dissipative quantum
dynamics.

All the above implies that the dynamics of the phases in the leads
$\chi_{\scriptscriptstyle R}$ and $\chi_{\scriptscriptstyle L}$
interacting with the heat bath differs from that of the internal
phases, i.e. the phases on superconducting islands $\chi_{ij}$.
Namely, whereas the phases in the leads are to be treated as
classical dynamic variables satisfying the Langevin stochastic
equation, the phases in the internal superconducting islands are the
quantum-mechanical variables with the dynamics described by the
quantum-mechanical Hamiltonian $H_0$. Moreover, the phases in the
leads and the intrinsic phases interact through the Josephson
coupling terms [the last two terms in the array Hamiltonian
(\ref{Hamilton-predv})]. Now, shifting all the phases over
$(\chi_{\scriptscriptstyle L}+\chi_{\scriptscriptstyle R})/2$, i.e
replacing $\chi_{ij}\rightarrow \chi_{ij}-(\chi_{\scriptscriptstyle
L}+\chi_{\scriptscriptstyle R})/2$, we obtain:
\begin{widetext}
\begin{eqnarray}\nonumber
H&=&H_0+H_{bath}+H_{int}\{\chi_{\scriptscriptstyle
L}-\chi_{\scriptscriptstyle R}\}+
\frac{\hbar^2}{8E_c}\sum_{j=1}^M(\dot\chi_{\scriptscriptstyle R}
-\dot\chi_{\scriptscriptstyle
L}-\dot{\chi}_{1j}+\dot{\chi}_{Nj})^2+\frac{\hbar^2}{8E_c}\sum_{j=1}^M(
\dot{\chi}_{1j}-\dot{\chi}_{Nj})^2\\&+&
\frac{\hbar^2}{2}\left(\frac{1}{E_c}+\frac{1}{E_{c0}}\right)\sum_{j=1}^M(\dot{\phi}_j)^2-2E_{\scriptscriptstyle
J}\sum_{j=1}^M\cos\phi_j\cos\left[ \frac{\chi_{\scriptscriptstyle
R}-\chi_{\scriptscriptstyle
L}-\chi_{1j}(t)+\chi_{Nj}(t)}{2}\right]\,, 
\label{Hamiltonian}
\end{eqnarray}
\end{widetext}
where a new \textit{independent} variable
$\phi_j=(\chi_{1j}+\chi_{Nj})/2$ had been introduced. The dc
Josephson current through the array is calculated as $I_s=\langle
\partial H/\partial [\chi_{\scriptscriptstyle R}-\chi_{\scriptscriptstyle
L}]\rangle$. Considering $\chi_{\scriptscriptstyle
R}-\chi_{\scriptscriptstyle L}$ as a parameter, one can apply the
Hellmann-Feynman theorem~\cite{Feynman} and reduce this
to~\cite{Matveev,AverinLikharev}
\begin{equation}
I_s(V)=\bigg\langle\frac{\partial \langle
H\rangle_{\phi_j}}{\partial[\chi_{\scriptscriptstyle
R}-\chi_{\scriptscriptstyle
L}]}\bigg\rangle_{\chi_{\scriptscriptstyle
R}-\chi_{\scriptscriptstyle L};\chi_{ij}}~. \label{Feynman}
\end{equation}

The term in the total Hamiltonian depending explicitly on the
variables $\phi_j$ is
\begin{widetext}
\begin{eqnarray}
\tilde{H}\{\phi_j\}=\frac{\hbar^2}{2}\bigg(\frac{1}{E_c}+\frac{1}{E_{c0}}\bigg)\sum_{j=1}^M(\dot{\phi}_j(t))^2
-2E_{\scriptscriptstyle J}\sum_{j=1}^M\cos \phi_j
 \times \cos\left[
\frac{\chi_{\scriptscriptstyle R}-\chi_{\scriptscriptstyle
L}-\chi_{1j}(t)+\chi_{Nj}(t)}{2}\right]. \label{HamiltonianPhi}
\end{eqnarray}
\end{widetext}
As a first step, we perform averaging over $\phi_j$. In the
insulator regime, which we address here, $E_c, E_{c0}\gg
E_{\scriptscriptstyle J}$, and the averaging procedure is carried
out by making use of the perturbation theory with respect to
$E_{\scriptscriptstyle J}/E_c$ [the last term in Eq.\,(\ref{HamiltonianPhi})]. 
Such a procedure is similar to the one used
in the case of the Cooper pair two-junctions
transistor~\cite{Matveev,Zorin}. In the zero order perturbation
theory $\langle \cos\phi_j\rangle_{\phi_j}=0$. Indeed, $\langle
\cos\phi_j\rangle_{\phi_j}~\propto~\langle n|\cos\phi_j|n
\rangle=0$, where  $|n\rangle=\prod_{j=1}^M
(1/\sqrt{2\pi})\exp(i\phi_j n)$ are the wave functions corresponding
to the first term in the Hamiltonian (\ref{HamiltonianPhi}), i.e.
the wave functions in a zero order of the perturbation theory.
The wave functions in the first order of the perturbation theory in
$E_{\scriptscriptstyle J}/E_c$ become \cite{Landau}
\begin{widetext}
\begin{equation}
\Psi_n=|n\rangle-2E_{\scriptscriptstyle J}\cos\left[
\frac{\chi_{\scriptscriptstyle R}-\chi_{\scriptscriptstyle
L}-\chi_{1j}(t)+\chi_{Nj}(t)}{2}\right] \sum_{j=1}^M \sum_m
\frac{\langle n|\cos
\phi_j|m\rangle}{E^{(0)}_n-E^{(0)}_m}|m\rangle~~,
\end{equation}
where
 \begin{equation}\nonumber
   E^{(0)}_n=\frac{E_cE_{c0}n^2}{2(E_c+E_{c0})}
 \end{equation}
are the energy levels obtained in a zero order of the perturbation
theory. Therefore, in the first order one finds
\begin{eqnarray}\nonumber
\big\langle \Psi_n|\cos\left[ \frac{\chi_{\scriptscriptstyle
R}-\chi_{\scriptscriptstyle L}-\chi_{1j}(t)+\chi_{Nj}(t)}{2}
\right]\cos \phi_j |\Psi_n \big\rangle_{\rho_T(\phi_j)}&=&
-\frac{E_{\scriptscriptstyle
J}(E_c+E_{c0})}{E_cE_{c0}}\sum_{n=-\infty}^{\infty}\frac{e^{-\frac{E_c n^2}{4k_BT}}}{(n^2-1/4)} \times\\
&~&\times \cos^2\left[ \frac{\chi_{\scriptscriptstyle
R}-\chi_{\scriptscriptstyle L}-\chi_{1j}(t)+\chi_{Nj}(t)}{2}
\right]\, . \label{averagephi}
\end{eqnarray}
Here, we used that for finite temperatures the quantum-mechanical
distribution of $\phi_j$ is determined by the equilibrium density
matrix corresponding to a first term in the Hamiltonian
(\ref{HamiltonianPhi}), i.e.
\begin{equation}\nonumber
\rho_{\scriptscriptstyle T}(\phi_j)~\simeq~\sum_n e^{-\frac{E_c
n^2}{4k_BT}}|n\rangle\langle n|\, .
\end{equation}
Further, to deal with less awkward formulas, we consider the case
$C\gg C_0$ (the general situation with the arbitrary relation
between $C$ and $C_0$ is straightforwardly recovered as needed) and
introduce the parameter
$$\alpha=\sum_{n=-\infty}^{\infty} \frac{e^{-\frac{E_c
n^2}{4k_BT}}}{(2n^2-1/2)}~.$$
The effective Hamiltonian $H_{eff}$ depending on the variables
$\chi_{\scriptscriptstyle R}-\chi_{\scriptscriptstyle L}$ and
$\chi_{ij}$ assumes the form:
\begin{equation}
H_{eff}=H_0+H_{bath}+H_{int}\{\chi_{\scriptscriptstyle
R}-\chi_{\scriptscriptstyle L} \}+H^{\star}\, ,
\end{equation}
where
\begin{equation} H^{\star}=\sum_{j=1}^M \frac{2\alpha
E_{\scriptscriptstyle J}}{E_c}E_{\scriptscriptstyle J} \cos\left[
\chi_{\scriptscriptstyle R}-\chi_{\scriptscriptstyle
L}-\chi_{1j}(t)+\chi_{Nj}(t)\right]\, , \label{HamiltonianInter}
\end{equation}
and, accordingly, the dc component of the \textit{non-dissipative}
Cooper pair Josephson current across the array is
\begin{equation}
I_s(V)=\bigg\langle\frac{\partial
H^{\star}}{\partial[\chi_{\scriptscriptstyle
R}-\chi_{\scriptscriptstyle L}]}\bigg\rangle=I_cM\frac{2\alpha
E_{\scriptscriptstyle J}}{E_c} \lim_{\tau \rightarrow \infty}
\frac{1}{\tau}\int_0^\tau dt \bigg\langle \sin\left[
\chi_{\scriptscriptstyle R}-\chi_{\scriptscriptstyle
L}-\chi_{1j}(t)+\chi_{Nj}(t)
 \right]\bigg\rangle~,
 \label{current-2}
\end{equation}
where, the brackets $\langle ...\rangle$ stand for a quantum
mechanical averaging over phases of internal junctions,
$\chi_{ij}(t)$.

The phase difference in the leads is fixed by the applied voltage
bias, $\chi_{\scriptscriptstyle R}-\chi_{\scriptscriptstyle
L}=2eVt/\hbar$, giving rise to a steady non-dissipative current.  We
will take into account the interaction with the thermal bath in the
leads by adding the Langevin thermal force $\psi(t)$ generating
phase fluctuations in the leads. This means that the effective
constraint on the phase on the leads can be written as:
\begin{equation}
\chi_{\scriptscriptstyle R}-\chi_{\scriptscriptstyle
L}=2eVt/\hbar+\psi(t)~, \label{classicalphases}
\end{equation}
where the phase correlation function in the leads $K_{leads}$ is
determined by the classical Nyquist noise in an external
resistance~\cite{Ingold,Shoenrev,KovFistUst}:
\begin{equation}
K_{leads}\equiv\langle\exp[\psi(t)-\psi(0)]\rangle_{\scriptscriptstyle
T}=\exp
\bigg\{\int_{-\infty}^{\infty}\frac{d\omega}{\omega}\frac{R_{ext}}{R_{\scriptscriptstyle
CP}[(\omega R_{ext}C)^2+1]}\bigg[\coth \bigg(\frac{\hbar
\omega}{2k_B T}\bigg)[\cos(\omega t)-1]-i\sin(\omega t)\bigg ]
\bigg\}\, . \label{k-leads}
\end{equation}
Thus the correlation function $K_{leads}$ determines the interaction
of the JJA with the thermal bath, and the energy relaxation takes
place only in the leads, but not inside the array
($\langle...\rangle_{\scriptscriptstyle T}$ stands for thermodynamic
average). Shifting all internal phases $\chi_{ij}-\chi_{kl}$ as
$\chi_{ij}-\chi_{kl}~=~\frac{2eV_{ij-kl}}{\hbar}t+\tilde{\chi}_{ij}-\tilde{\chi}_{kl}$,
we bring $H_0$ to the form (we omit the tilde-sign):
\begin{equation}
H_0=\sum_{\langle
ij,kl\rangle}\Bigl[\frac{\hbar^2}{4E_c}(\dot{\chi}_{ij}-\dot{\chi}_{kl})^2-
\tilde{E}_J\cos\Bigl(\frac{2eV_{ij-kl}t}{\hbar}+\chi_{ij}-\chi_{kl}\Bigr)\Bigr]
+\sum_{ij}\frac{\hbar^2}{4E_{c0}}\dot{\chi}_{ij}^2~, \label{ham-new}
\end{equation}
where $V_{ij-kl}(t)$ are voltage drops between the adjacent islands.
Plugging (\ref{classicalphases}) into (\ref{HamiltonianInter}) we
find that the term $H^{\star}$ in the Hamiltonian can be viewed as a
time-dependent perturbation (the ac Josephson current) oscillating
with the frequency $\omega~=2e(V_{1j}+V_{Nj})/\hbar$:
\begin{equation} H^{\star}=\frac{2\alpha E_{\scriptscriptstyle J}}{E_c}E_{\scriptscriptstyle J} \sum_{j=1}^M
\cos\left[\frac{2e(V_{1j}+V_{Nj})t}{\hbar}
+\psi(t)-\chi_{1j}(t)+\chi_{Nj}(t)\right] \label{HamiltonianInter-2}
\end{equation}
\end{widetext}
and that the average value of the dc current can be considered as
calculated in the first order with respect to this perturbation
\cite{FickSau,Ingold2}. Here, $V_{1j}=V_1$ and
$V_{Nj}=V_{\scriptscriptstyle N}$ are the dc voltage drops between
the left lead and first row of islands, and the $N$-th row of
islands and the right lead, accordingly. Following the general
recipe \cite{FickSau,Ingold2,Cohen}, we carry out the averaging in
Eq.~(\ref{current-2}) with the help of the nonequilibrium density
matrix $\rho(t)$, which satisfies the equation
\begin{equation}
\dot{\rho}(t)=-i\hat L(t)\rho(t)~,
 \label{densitymatrix-gen}
\end{equation}%
where the Liouville operator is determined as
\begin{equation}
\hat L \hat X=\frac{1}{\hbar}[H_0+H_{bath}+H_{int}+H^{\star},\hat X]\,. 
\label{Liouville}
\end{equation}
Solving the Eq.\,(\ref{Liouville}) up to the first order in
$H^{\star}$ we obtain
\begin{equation}
\rho(t)=\rho_\beta-\frac{i}{\hbar}\int_0^t ds e^{-\hat
L_0(t-s)}[H^{\star},\rho_\beta]~, \label{densitymatrix}
\end{equation}%
where $\rho_\beta$ is the equilibrium density matrix corresponding
to the Hamiltonian $H_0+H_{bath}+H_{int}$, the Liouville operator
$\hat L_0$ is determined as $\hat L_0 \hat
X=(1/\hbar)[H_0+H_{bath}+H_{int},\hat X]$. The expression for the
average dc Cooper pair current assumes the form:
\begin{widetext}
\begin{equation}
I_s(V)=MI_c\frac{4E_{\scriptscriptstyle J}}{\hbar}\frac{\alpha^2
E_{\scriptscriptstyle J}^2}{ E_c^2}\lim_{\tau \rightarrow \infty}
\frac{1}{\tau}\int_0^\tau dt\int_0^t ds
\sin\bigg[\frac{2eV}{\hbar}(t-s)\bigg]\bigg\langle [\hat F_1(t-s),
\hat F_1]+[\hat F_2(t-s), \hat F_2]\bigg\rangle_{H_0,\psi}~,
 \label{current-3}
\end{equation}
where $[\hat F_{1,2}(t-s), \hat F_{1,2}]$ is the commutator of the
operator $\hat F_{1,2}$ and the corresponding Heisenberg operator
$\hat F_{1,2}(t-s)$. The functions $F_{1,2}$ are determined as
\begin{equation}
F_{1}=\cos \big[\psi-\chi_{1j}+\chi_{Nj}\big],~\hspace{1cm} F_{2}=\sin
\big[\psi-\chi_{1j}+\chi_{Nj}\big]\, .
\end{equation}
We define the time-dependent correlation function $K_{tot}(t)$ of a
whole system (including leads) as
\begin{eqnarray}
K_{tot}(t)=\bigg\langle \exp\bigg\{i\big[\psi(t)-\psi(0)-
\chi_{1j}(t)+\chi_{1j}(0)+\chi_{Nj}(t)-\chi_{Nj}(0)\big]\bigg\}\bigg\rangle_{H_0,\psi}~.
 \label{totalKorfunction}
\end{eqnarray}
Using the property of average values \cite{Ingold2} $ \langle
F(t)G(0)\rangle_{\rho_\beta}=\langle F(0)G(-t)\rangle_{\rho_\beta} $
and expressing all the commutators through $K_{tot}(t)$ we arrive at
the general equation for the dc current
\begin{eqnarray}
I_s(V)=8\alpha^2 MI_c\frac{E_{\scriptscriptstyle
J}}{\hbar}\frac{E_{\scriptscriptstyle J}^2}{
E_c^2}\int_{-\infty}^{\infty}dt \sin \bigg[
\frac{2e(V_1+V_{\scriptscriptstyle N})}{\hbar}t\bigg ] \Im m
[K_{tot}(t)]~. \label{current-4}
\end{eqnarray}

Since $H_0$ does not contain the variable $\psi$ and $H_{int}$, in
its turn, does not depend on the intrinsic variables $\chi_{ij}$,
the correlation function (\ref{totalKorfunction}) factorizes:
\begin{equation}
K_{tot}(t)=K_{leads}(t)K(t) \label{Factorization}\, ,
\end{equation}
where the correlation function of the phase noise in leads is
determined by Eq.(\ref{k-leads}) reflecting as we have already
mentioned the interaction with the thermal bath (see (\ref{Bath})
and (\ref{InteractionBath})) and
\begin{equation}
K(t)=\langle \exp
i\left[\chi_{1j}(t)-\chi_{1j}(0)-\chi_{Nj}(t)+\chi_{Nj}(0)
\right]\rangle_{H_0}~,
\label{corrfunct}
\end{equation}
is the time-dependent correlation function of the intrinsic part of
the system with $H_0$ defined in (\ref{ham-new}).
\end{widetext}
Note here that in a two-junction system (a single Cooper-pair
transistor) where $\chi_{1j}=\chi_{Nj}$ and $K(t)\equiv 1$, the
current-voltage characteristics $I(V)$ is determined by the external
resistance \cite{Zorin}.  We calculate the $I$-$V$ curve for the
current flowing through a single Cooper-pair transistor as an
example of an application of the general formalism. If the external
resistance $R_{ext}<R_{\scriptscriptstyle CP}$, the current displays
a peak in the low-voltage region (this state is often referred to as
the ``pseudo-superconducting state"):
\begin{equation}
I(V)~\simeq~MI_cE_{\scriptscriptstyle
J}\bigg[\frac{E_{\scriptscriptstyle J}}{E_c}\bigg]^2
\frac{R_{ext}}{R_{\scriptscriptstyle CP}}
\frac{eV}{e^2V^2+(k_BTR_{ext}/R_{\scriptscriptstyle CP})^2}\,.
\label{I-Vcurve2JJs}
\end{equation}
Proportionality of $I(V)$ to $E_{\scriptscriptstyle J}^4$ indicates
the Cooper-pair cotunneling type of transport. Note, that the
Cooper-pair current in the two-junction system \emph{does not }
display Coulomb blockade effect when the external resistance
$R_{ext}<R_{\scriptscriptstyle CP}$. In the opposite case,
$R_{ext}>R_{\scriptscriptstyle CP}$, the Eq.\,(\ref{current-4}) with
$K(t)=1$ yields the Gaussian $I(V)$ dependence
$I\propto\exp[-(eV-E_c)^2/(k_BTE_c)]$ (see
Ref.\,[\onlinecite{Ingold}]).

Now we discuss the general case of a large size JJA with the number
of Josephson junctions larger than two.  We consider the case of the
small external resistance, $R_{ext}<R_{\scriptscriptstyle CP}$,
which is the most frequent experimental situation.
Expression (\ref{current-4}) shows that the dc current depends
explicitly upon the voltage drops $V_1$ and $V_{\scriptscriptstyle
N}$ on the leftmost and rightmost junctions, while the voltage drops
$V_{ij-kl}$ on the internal parts of the array come in through the
correlation function $K(t)$.  To evaluate the voltage distribution
along the system, we notice that in the insulating domain,
$\tilde{E}_{\scriptscriptstyle J}\ll E_c$, the correlation function
$K(t)$ oscillates with the high frequency determined by the
macroscopic collective Coulomb gap $\Delta_c\gg k_BT\gg E_c$ given
by Eq.\,(\ref{Deltac}), and that in the time interval, from which
the main contribution is coming from, $K_{leads}\approx 1$ (see two
next Sections), giving rise to the exponentially low conductance
$G(T)~\propto~G_0 e^{-\Delta_c/(k_BT)}$ (where $G_0$ is the
non-activated factor in the conductance characterizing an individual
junction), i.e.
\begin{equation}
I~\simeq~G(T)(E_{\scriptscriptstyle
J}/E_c)^4(V_1+V_{\scriptscriptstyle N})~ \label{Current-BJ}.
\end{equation}
Then the \emph{dc} current passing through an internal junction can
be estimated as [in the first order approximation with respect to
the time-dependent terms $\tilde{E}_{\scriptscriptstyle
J}\cos[(2eV_{ij-kl}t/\hbar)+\chi_{ij}-\chi_{kl}]$ in the Hamiltonian
$H_0$ of (\ref{ham-new})]:
\begin{equation}
I^{(int)}_{dc}~\simeq~G_0V_{ij-kl} (\tilde{E}_{\scriptscriptstyle
J}/E_c)^2 \label{Current-Intr}
\end{equation}
[cf Eqs. (\ref{densitymatrix})-(\ref{current-4})]. Utilizing the dc
current conservation law (the Kirchhoff´s law), i.e. the fact that
$I=I^{(int)}_{dc}$, we find:
\begin{equation}
V_1+V_{\scriptscriptstyle
N}~\simeq~\frac{V}{1+\big[NE^4_{\scriptscriptstyle
J}/\big(\tilde{E}_{\scriptscriptstyle
J}E_c\big)^2\big]\big[{G(T)/G_0}\big]}~,\label{Voltagedrop}
\end{equation}
and, therefore, as long as
\begin{equation}
N\frac{E^4_{\scriptscriptstyle J}} {(\tilde{E}_{\scriptscriptstyle
J}E_c)^2} \frac{G(T)}{G_0}\ll 1\, ,~\label{Condition}
\end{equation}
the highly inhomogeneous voltage distribution takes place, i.e.
almost all the applied dc bias $V$ drops on the first and the $N$-th
rows of junctions, $V_1+V_{\scriptscriptstyle N}~\approx~V$. In this
regime, the dc current $I~\simeq~G(T)(E_{\scriptscriptstyle
J}/E_c)^4V$ flowing through the array \emph{seems} not to depend on
the Josephson coupling of intrinsic Josephson junctions
$\tilde{E}_{\scriptscriptstyle J}$. However, as
$\tilde{E}_{\scriptscriptstyle J}$ becomes too small and/or, talking
about the 2D case, the number $N$ of junctions is growing too large,
the condition (\ref{Condition}) breaks down. 
This gives rise to the even distribution of the total
voltage drop  along the whole array.
Yet due to extremely small values of $G(T)$ in the insulating
regime, there exists a wide range of parameters where the highly
inhomogeneous voltage distribution giving rise to the synchronized
collective behavior is realized.

The above estimates suggest the following simple picture of the
macroscopic Coulomb blockade governing the Cooper pair insulator
dynamics: the applied voltage distributes mostly between the
leftmost and rightmost rows of junctions, while the internal part of
an array acts as a coherent superconducting island providing the
macroscopic Coulomb barrier $\Delta_c\simeq(2e)^2/2C_{tot}$, where
$C_{tot}$ is a total capacitance of the array.

\section{Transport in JJAs: the insulating regime}
In this Section we calculate the current defined by Eqs.
(\ref{current-4}) - (\ref{corrfunct}), in the {\it insulating
state}, $E_c\gg E_{\scriptscriptstyle J}$, at moderately low
temperatures $\Delta_c>k_B T\gtrsim E_c$.

To begin with, we find the correlation function $K(t)$ for a
simplest 1D case.  In the first approximation we put
$\tilde{E}_{\scriptscriptstyle J}=0$, and, to simplify the
notations, assume $E_{c0}=\infty$ (the generalization to the finite
$E_{c0}$ case is almost straightforward). Then the Hamiltonian
$H_0=\sum_i H_i$, where $H_i$ are the Hamiltonians of individual
Josephson junctions. Using the quantum-mechanical definition of
$K(t)$ we obtain\cite{FickSau}
$$
K(t)=\bigg[A\sum_{n,m}e^{it[E_m-E_n]-\frac{E_n}{k_B T}}|\langle
n|e^{i\varphi}|m\rangle|^2\bigg]^N~,
$$
where $E_n$'s are the energy levels of a single junction. These
energy levels are determined by the charge energy $E_c$ as $E_n=(E_c
n^2)/2$, $\langle n|e^{i\varphi}|m\rangle$ are the matrix elements
of the operator $e^{i\varphi}$ between the $n$-th and $m$-th states,
and $A=[\sum_n\exp(-E_n/k_BT)]^{-1}$ is the normalization
coefficient (which cancel out from the final expression for $K(t)$).
Therefore, the quantum mechanical dynamics of a single junction can
be mapped onto a well studied behavior of quantum rotator which has
the matrix elements for the operator $e^{i\varphi}$ between the
states $n$ and $n+1$ only. We find
$$
K(t)=e^{iNE_ct/2}\bigg[A\sum_n
e^{inE_ct-\frac{E_cn^2}{2k_BT}}\bigg]^N\, .
$$
At $E_c < k_B T$ we replace the sum by the integral
$$
K(t)=e^{iNE_ct/2}\bigg[A\int dx \exp \{iE_ct x-E_cx^2/(2k_BT)\}
\bigg]^N,
$$
and arrive at
\begin{equation}
K(t)=e^{iNE_ct/2-NE_ck_BTt^2/2}~~. \label{corrfunction}
\end{equation}

In a general 1D case with the finite value of $E_{c0}$ the time
dependence of the correlation function $K(t)$ preserves its form
(\ref{corrfunction}) but the quantity $NE_c$ has to be replaced by a
more general expression determined by the full capacitance matrix of
the array. A crucial conditions allowing to obtain this result are
the bilinear form of the Hamiltonian $H_0$ in the momentum
representation, and the factorization of the wave functions:
$|n\rangle_{H_0}=\prod_{k=1}^N(1/\sqrt{2\pi})\exp(i\phi_k n_j)$.

In the two-dimensional situation the calculations are more involved
and the correlation function of the array, $K(t)$, is derived as an
analytical continuation of the quantity $K(\tau)$, where $\tau$ is
the imaginary time,
\begin{widetext}
\begin{eqnarray}\nonumber
K(\tau)=\int D[\chi_{ij}]e^{
i[\chi_{1j}(\tau)-\chi_{1j}(0)-\chi_{Nj}(\tau)+\chi_{Nj}(0)]}
\exp\bigg(-\frac{\hbar}{4}\int_0^{\hbar/(k_BT)}d\tilde\tau \bigg[
\sum_{\langle
ij,kl\rangle}\frac{[\dot{\chi}_{ij}(\tilde\tau)-\dot{\chi}_{kl}(\tilde\tau)]^2}{E_c}\\
-\tilde{E}_{\scriptscriptstyle
J}\cos[\chi_{ij}(\tilde\tau)-\chi_{kl}(\tilde\tau)]
-\sum_{ij}\frac{[\dot{\chi}_{ij}(\tilde\tau)
]^2}{E_{c0}}\bigg]\bigg)\,. \label{korrfunctCalc}
\end{eqnarray}
Note that the correlation function $K(t)$, determining the ac
synchronization between the external leftmost, 1-st, and
rightmost, $N$-th, Josephson junctions contacting with the left
and right leads respectively, \emph{is not zero} even in the
zero-approximation, $\tilde{E}_{\scriptscriptstyle J}=0$, with
respect to the \emph{intrinsic} Josephson coupling
$\tilde{E}_{\scriptscriptstyle J}$. The phases $\chi_{ij}$ are
written as ${\chi}_{ij}+2\pi M_{ij}(k_B T \tau/\hbar)$, where
${\chi}_{ij}$ is a periodic function on the interval
$0<\tau<\hbar/(k_B T)$, and $M_{ij}$ are the winding numbers.

In the high temperature regime $k_BT\gg E_c $ we can neglect all
nonzero winding numbers. Indeed, the nonzero winding numbers
contribution to the $K(\tau)$ can be estimated as:
$$
K^{(M)}(\tau)~\simeq~\exp\{(M_{1j}-M_{Nj})\frac{2\pi k_B
T\tau}{\hbar}-\sum_{\langle ij,kl\rangle}\frac{(\pi^2
k_BT)}{E_c}[M_{ij}-M_{kl}]^2 \}~.
$$
Therefore, on the time scale $\tau\ll \hbar/E_c$ the contribution of
nonzero windings numbers is small. Since the characteristic time
$\tau~\simeq~\hbar/(k_B T)$ in the integral over time in the 
Eq.\,(\ref{current-4}), we neglect all nonzero winding numbers in the
limit $E_c/k_BT ~\ll ~1$. The nonzero winding numbers become
important at low temperatures, $k_BT\ll E_c$.

Next, we expand the periodic phases ${\chi}_{ij}(\tau)$ over the
Matsubara frequencies (see also Appendix):
\begin{equation}
\chi_{ij}(\tau)=\sum_{\omega_n=2\pi k_B Tn/\hbar}e^{i\omega_n
\tau}\chi_{ij}(\omega_n)\,, \label{matsubara}
\end{equation}
and change the variables in the integrals over
$\chi_{ij}(\omega_n)$:
\begin{equation}
\chi_{ij}(\omega_n)=x_{ij}(2E_c k_B
T/(\hbar\omega_n)^2)[\exp(-i\omega_n \tau)-1]\,.
\label{chvariab}
\end{equation}
Substituting (\ref{matsubara}) and (\ref{chvariab}) into
(\ref{korrfunctCalc}) yields the correlation function $K(t)$ in the
following form
\begin{equation}
K(\tau)=\int D[x_{ij}]\exp \bigg\{\sum_{\omega_n}\frac{16E_ck_BT
\sin^2(\omega_n\tau/2)}{\hbar^2\omega_n^2}
 \bigg[i(x_{1j}-x_{Nj}) -
\sum_{\langle ij,kl\rangle}\frac{1}{2}(x_{ij}-x_{kl})^2
-\sum_{ij}\frac{E_c x^2_{ij} }{2E_{c0}}\bigg]\bigg\}\,.
\label{korrfunctCalc2}
\end{equation}
\end{widetext}

The function $K(\tau)$ is the \emph{periodic} function of $\tau$
with the period $=\hbar/(k_BT)$. Therefore, it is enough to
calculate the sum over $\omega_n$ in (\ref{korrfunctCalc2}) for
$0<\tau<\hbar/(k_BT)$. In this range of $\tau$
$$
\sum_{\omega_n}\frac{16E_ck_BT\sin^2(\omega_n\tau/2)}{\hbar^2\omega_n^2}=
\frac{4E_c}{\hbar}\tau-\frac{4E_ck_BT}{\hbar^2}\tau^2~,
$$
and  one finally arrives at
\begin{equation}
K(\tau)= \exp{\bigg(\frac{4\Delta_ck_B
T\tau^2}{\hbar^2}-\frac{4\Delta_c\tau}{\hbar}\bigg)}\label{korrfunctlarge-imagT}~,
\end{equation}
where $\Delta_c$ is the \textit{macroscopic Coulomb gap} for the
Cooper pair propagation defined through the functional integral on
the lattice:
\begin{eqnarray}\nonumber
\exp(-\Delta_c/k_BT)= \int D[x_{ij}]\exp
\frac{E_c}{k_BT}\bigg[i(x_{1j}-x_{Nj})\\
-\sum_{\langle ij,kl\rangle}\frac{1}{2}(x_{ij}-x_{kl})^2
-\sum_{ij}\frac{E_c x^2_{ij} }{2E_{c0}}\bigg]\,.
 \label{Coulombgap}
\end{eqnarray}

The analytic continuation of the periodic function $K(\tau)$ to the
real time $t$, and the corresponding calculation of $\Im mK(t)$ is
to be carried out according to the general recipes of the
statistical physics~\cite{AGD}. First, we find the quantities
$K(\omega_n)$ determined by the Matsubara frequencies
($\omega_n=2\pi n(k_BT)/\hbar$):
\begin{equation}
K(\omega_n)=\int_0^{\hbar/(k_BT)}K(\tau)e^{i\omega_n \tau}d\tau~
\label{Fouriertransform}
\end{equation}
[we remind that $K(\tau)$ is the periodic function of $\tau$ on the
interval $(0, \hbar/(k_BT))$]. The next step is the analytic
continuation $i\omega_n \rightarrow \omega+i\delta$, giving rise to
the retarded correlation function $K^R_\omega$. Performing then the
inverse Fourier transformation, we get:
$$
K^R(t)=2\Im m K(t)=\int_{-\infty}^{\infty}K^R_\omega \exp{(-i\omega
t)} \frac{d\omega}{2\pi}\,.
$$

\begin{figure}[tbp]
\includegraphics[width=2.6in]{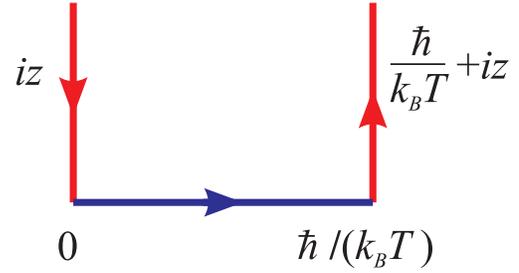}
\caption{The contour for calculation of the integral in
Eq.\,(\ref{Fouriertransform}).} \label{Contour}
\end{figure}

To carry out the analytic continuation we transform the integral in
(\ref{Fouriertransform}) from the real axis to the contour in the
complex plain, i.e. two lines $iz$ and $\hbar/(k_BT)+iz$, where $z$
runs first from $\infty$ to $0$ and then from $0$ to $\infty$ (see
Fig.\,\ref{Contour}) and find
\begin{equation}
K(\omega_n)=i\int_0^{\infty}dz
\bigg[K(iz)-K(iz+\frac{\hbar}{k_BT})\bigg]e^{-\omega_n z}\,.
\label{Fouriertransform-new}
\end{equation}
After that, we change $i\omega_n$ to $\omega+i\delta$, and
performing the inverse Fourier transform, we obtain \cite{comment4}
\begin{equation}
\Im m K(t)=-i\bigg[K(it)-K(it+\frac{\hbar}{k_BT})\bigg]\,,
\label{Continuation}
\end{equation}
and in the limit $E_c\ll k_BT\ll\Delta_c$ we finally arrive at
\begin{equation}
\Im m K(t)=\exp{\bigg(-\frac{4\Delta_ck_B
Tt^2}{\hbar^2}\bigg)}\sin\bigg(\frac{4\Delta_ct}{\hbar}\bigg)\label{korrfunctlargeT}\,.
\end{equation}
Calculating the integral over time in Eq.~(\ref{current-4}) we find
the $I$-$V$ dependence in the following form
\begin{equation}
I~\simeq~(V_1+V_{\scriptscriptstyle
N})\exp\bigg[-\frac{(2\Delta_c-e(V_1+V_{\scriptscriptstyle
N}))^2}{4k_BT\Delta_c}\bigg]\label{I-Vcurve}~.
\end{equation}
Substituting the expression for $V_1+V_{\scriptscriptstyle N}$
(\ref{Voltagedrop}) in (\ref{I-Vcurve}) we obtain the $I(V)$
characteristic of a large JJA in a form:
\begin{equation}
I\propto~\tilde{V}\exp\bigg[-\frac{(2\Delta_c-e\tilde{V})^2}{4k_BT\Delta_c}\bigg]\,,
\label{I-Vexact}
\end{equation}
with
$$
\tilde{V}=V\bigg(1+N\frac{E_{\scriptscriptstyle
J}^4}{(\tilde{E}_JE_c)^2}\frac{G(T)}{G_0}\bigg)^{-1}\,.
$$
Note that if we set $\tilde{E}_{\scriptscriptstyle J}=0$ in
Eq.~(\ref{I-Vexact}), then the current through the system is zero,
$I=0$, as it should.  On the other hand, if the condition
(\ref{Condition}) is satisfied, $V_1+V_{\scriptscriptstyle
N}\approx V$, and the $I$-$V$ curve assumes the simple final form
\begin{equation}
I~\simeq~V\exp\bigg[-\frac{(2\Delta_c-eV)^2}{4k_BT\Delta_c}\bigg]\,.
\label{I-Vcurve-Fin}
\end{equation}
The Gaussian formula (\ref{I-Vcurve-Fin}) was obtained earlier for a
single JJ incorporated in a circuit with the high resistance; the
corresponding peak in the $I$-$V$ curve was considered a
manifestation of the ``Coulomb blockade of Cooper-pair tunneling"
\cite{Shoenrev}. Experimentally such a peak has been observed in
Refs.~[\onlinecite{Kuzmin,Kuzmin2}]. In our case of large JJAs one
does not expect the similar Gaussian peak in the $I$-$V$ dependence.
The reason is that on approach of the bias $eV$ to $\Delta_c$, the
conductance $G(T)$ grows appreciably and the condition
(\ref{Condition}) breaks down.  The voltage distribution becomes
homogeneous and formula (\ref{I-Vcurve-Fin}) does not hold any more.

At low voltages, $eV\ll \Delta_c$, Eq.\,(\ref{I-Vcurve-Fin}) yields
the thermally activated behavior of the resistance:
\begin{equation}
R_{array}\propto\exp\bigg(\frac{\Delta_c}{k_B T}\bigg)\, ,
\label{activation}
\end{equation}
and therefore we identify the activation temperature $T_0$ of the
experiment with the macroscopic Coulomb blockade barrier
$\Delta_c/k_B$.  In order to carry out calculations in
Eq.~(\ref{Coulombgap}) and determine $\Delta_c$, we consider the
standard  ``spin-wave types" fluctuations:
$$x_{ij}=\int
d\vec{p}\exp{(i\vec{p}\vec{R}_{ij})}x(\vec{p})\,.
$$
Taking all the Gaussian integrals over $x(\vec{p})$ in
Eq.~(\ref{Coulombgap}) we obtain the expression for $\Delta_c$ in
the following form (the vector $\vec{L}$ is directed along the
current in  JJA):
\begin{equation}
\Delta_c=2E_c d^{n-2}\int \frac{d^n \vec{p}}{(2\pi)^n}\frac{\sin^2
\frac{\vec{p}\vec{L}}{2}}{[p^2+(1/\lambda_c)^2]}\,,
\label{DeltaC-gen}
 \end{equation}
where $\lambda_c~=d\sqrt{E_{c0}/E_c}$ is the correlation length in
the charge coupled tunnel junction arrays, and $n=1,2$ for the 1D
and 2D JJAs, correspondingly. For a large one-dimensional array,
$L\gg \lambda_c$,
\begin{equation}
\Delta_c=E_c\lambda_c/(2d) \,.
 \label{DisordLarge1}
\end{equation}
In the opposite limit of shorter one-dimensional arrays,
i.e. $L\ll \lambda_c$, the Coulomb gap increases with the array size $L$
linearly:
\begin{equation}
 \Delta_c=E_cL/(2d).
 \end{equation}
In two-dimensional junction arrays the Coulomb gap acquires the
logarithmic form:
\begin{equation}
\Delta_c= E_c \ln \frac{\min{\{\lambda_c,L}\}}{d}\,.
 \label{DisordAver2}
\end{equation}
This concludes the description of thermally activated behavior in
the temperature interval $E_c<k_BT<\Delta_c$.

\section{Transport in JJAs: the superinsulating regime}

Now we turn to low temperatures, $k_BT\ll E_c$, where all the
windings numbers $M_{ij}$ have to be taken into account.  In a
one-dimensional array the calculation of the correlation function
$K(t)$ is straightforward.  Consider, to be specific, the case where
the screening length is larger than the size of a system $L$, then:
$$
K(t)=e^{ \frac{iNE_ct}{2} }\bigg[A\sum_n e^{iE_c n
t-\frac{n^2E_c}{2k_BT} } \bigg]^N~,
$$
\begin{equation}
A=\bigg[\sum_n e^{-\frac{n^2E_c}{2k_BT} }\bigg]^{-1}~,
\label{corrfunctQL}
\end{equation}
and the values $n=0, \pm 1$ give the main contribution. As a result
we arrive at
\begin{equation}
K(t)=\exp\bigg\{i\Delta_c t+2Ne^{-E_c/(2k_BT)}[\cos(E_c t)-1]
\bigg\}\, . \label{corrfunctQLfin}
\end{equation}
Substituting (\ref{corrfunctQLfin}) into Eq. (\ref{current-4}) we
find the expression for the current as
\begin{equation}
I(V)~\propto\exp\bigg[-\frac{(eV-\Delta_c)^2 e^{E_c/2k_B
T}}{8E_c^2N}\bigg]~.\label{currentQL}
\end{equation}
This result holds in the temperature range $ E_c/\ln(N)<k_B T \ll
E_c$.

In the two-dimensional case, the contribution from the nonzero
windings numbers is analogous to the vortex contribution which
appears in the classical two-dimensional planar Heisenberg model (or
classical Josephson junction arrays below the BKT). We will follow
the procedure developed for calculation of the vortex contribution
to various correlation functions in Ref.\,[\onlinecite{Jose}].
Namely, it was shown that the coordinate dependent correlation
function
$$
g^p_{ (vortex) }(\textbf{r}-\textbf{r}^{\prime} )=\langle\exp \big\{
ip\big[\chi(\textbf{r})-\chi(\textbf{r}^{\prime})\big]
\big\}\rangle~,
$$
where $\textbf{r}$ and $\textbf{r}^\prime$ are the two points on the
2D lattice, can be expressed as
$$
g^p_{(vortex)}(\textbf{r}-\textbf{r}^{\prime})~\simeq~
\exp\left(-\frac{\pi}{4}p^2
\xi\ln\frac{|\textbf{r}-\textbf{r}^{\prime}|}{d}\right),
$$
where $\xi=\sum_{r_0} r_0^2\langle m(0)m(r_0)\rangle$ is the space
correlation function of vortex (charge)-antivortex (anticharge)
pairs, diverging near the binding-unbinding transition temperature.
Using this result and taking into account the corresponding mapping
$p=\tau E_c/\hbar$, we find
\begin{equation}
K(t)= \exp{\bigg(-\frac{\Delta_c E_c \xi
t^2}{\hbar^2}-i\frac{2\Delta_ct}{\hbar}\bigg)}\label{korrfunctsmallT}~.
\end{equation}
Plugging (\ref{korrfunctsmallT}) into (\ref{current-4}) and
calculating integral over time, we obtain, at low voltages, the
following expression for the resistance of the array:
\begin{equation}
R_{array}~\propto~\exp\bigg(\frac{\Delta_c}{E_c\xi}\bigg)
\label{activation-1}\,.
\end{equation}

At low temperatures, $k_BT\ll E_c$ the concentration of the
charge-anticharge pairs is small and accordingly,
$\xi=\text{const}\cdot\exp[-E_c/(k_B T)]$ (see~Ref.\,[\onlinecite{Jose}]). This
gives double-exponential behavior of the resistance
\begin{equation}
R\propto\exp\left[\frac{\Delta_c}{E_c}\exp\left(\frac{E_c}{k_BT}\right)\right]\,
\end{equation}
in the superinsulating regime.  We would like to emphasize here that
the double-exponential temperature dependence favors enormously the
fulfilling the condition~(\ref{Condition}) for the inhomogeneous
distribution of the voltage drop, which ensures the validity of our
approach.
\section{A qualitative picture}

\begin{figure}[tbh]
\includegraphics[width=3.2in]{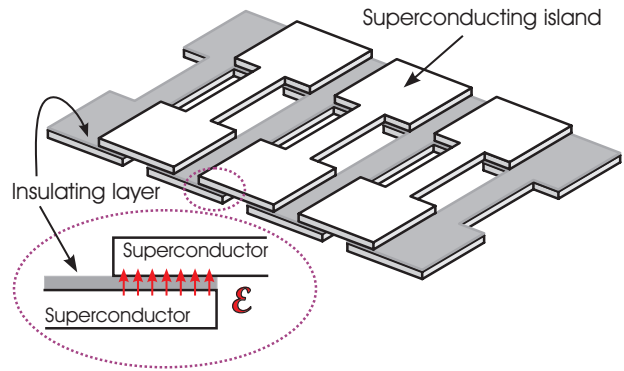}
\caption{Distibution of the electric field in one-dimensional array
of superconducting islands connected by two Josephson junctions to
neighbors corresponding to experimental system of Ref.\,[\onlinecite{Haviland1}].
} \label{ElctricField}
\end{figure}

To gain physical insight in the transport phenomena near SIT of the
large one- and two-dimensional Josephson junction arrays and films,
let us discuss the distribution of the electric field in the
experimental systems in question. Consider first one- and
two-dimensional
JJAs~\cite{Mooij1,Mooij2,Mooij3,Mooij4,Haviland1,Haviland2,Tighe,Japan1,
Japan2}. The arrays are comprised of overlapping superconducting
platelets (islands) separated by thin oxide layers (see
Fig.\,\ref{ElctricField}). The related junction capacitance, $C$,
well exceeds the capacitance of each constitutive island to the
ground, $C_0$. Thus, the total capacitance of the JJA is determined
by the capacitance of a junction $C$.   If now we place a charge in
the array, the induced electric field will remain within the array
plane. In other words, one- and two-dimensional arrays can be viewed
as systems with the anomalously large dielectric constant
$\varepsilon\simeq C/C_0$. Accordingly, in 1D arrays charges
interact linearly over distances $l<\lambda _c$, and in 2D arrays
the charges interact logarithmically over scales
$l<\lambda _c$~Ref.\,[\onlinecite{Mooij4}].

\begin{figure}[tbh]
\includegraphics[width=3.2in]{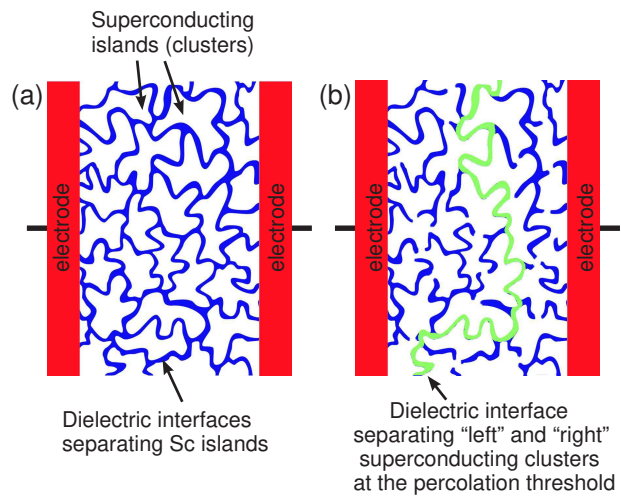}
\caption{Developing of the large capacitance in the disordered film
on approach to the superconductor-to-insulator transition from the
insulating side. (a) Superconducting clusters separated by
dielectric interfaces. (b) The same picture at the immediate
vicinity of the SIT.  Cutting the last interface separating the left
and right superconducting clusters (green line) gives rise to
percolation over superconducting areas from the left electrode to
the right electrode, i.e. to the superconducting state. }
\label{ElctricField-film}
\end{figure}

Turning to disordered superconducting films and granular
superconductors near the SIT, we recall that in the two phase system
in the vicinity of the percolation transition between the conducting
and insulating phases, the dielectric constant
diverges~\cite{Mott68,Dubrov}.  The origin of this divergence can be
understood from the simple picture of the percolation transition,
see Fig.~\ref{ElctricField-film}.  The white spots represent
superconducting clusters in the insulating sea (dark blue). The
capacitance between the two adjacent clusters is proportional to the
length of the insulating layer separating them.  Upon approaching
the transition from the insulating side of the SIT (see
Fig.\,\ref{ElctricField-film}b), the length of this layer diverges
infinitely. It results in the divergent growth of the effective
capacitance of the system, implying the divergence of the dielectric
constant.  Since the SIT in disordered films and granular
superconductors is supposed to be of the percolative
nature~\cite{LarkinOvchin71n1,IoffeLarkin,MaLee,Imry}, one expects
that these systems possess anomalously large $\varepsilon$ near the
transition.  Recently the enhanced dielectric constant was indeed
observed near the SIT in ultrathin amorphous beryllium
films~\cite{BeDSIT}. We therefore can conclude that superconducting
films near the SIT exhibit two dimensional behavior with respect to
Coulomb interaction. Note that from the viewpoint of the Coulomb
interaction between charges, the 2D JJAs and disordered films near
the SIT are alike: what matters is the logarithmic interaction
between the charges, while the differences in internal structure
between the systems in question are irrelevant (at least in the
absence of the magnetic field).

The two-dimensional character of the Coulomb interaction has
important implications for the systems with the size less then the
electrostatic screening length $\lambda_c$. Unbinding of
logarithmically interacting topological excitations gives rise to
the celebrated universal Berezinskii-Kosterletz-Thouless (BKT) phase
transition first introduced in the context of vortices in XY
-magnets and extended then to vortex-antivortex pairs in superfluid
and superconducting films and Josephson junction
arrays~\cite{BKT-1,BKT-HalpNelson,BKT-Don,BKT-Minn}.

On the superconducting side of the superconductor-insulator
transition logarithmic interaction between vortices gives rise to
BKT transition separating the superconducting low-temperature phase,
where vortices and antivortices are bound in pairs, from the high
temperature phase with free vortices. In the high-temperature domain
the free motion of vortices breaks down the phase coherence and a
superconductor falls into the resistive state with the resistance
much less than that in the normal state. At $T=T_{\scriptscriptstyle
BKT}\simeq E_{\scriptscriptstyle J}/k_{\scriptscriptstyle B}$ the
phase coherence restores and the 2D array or film becomes
superconducting.  On the insulating side the film experiences
\textit{charge} binding-unbinding transition at the temperature
$T=T_{\scriptscriptstyle SI}\simeq
E_c/k_B$~[\onlinecite{Sugahara,Widom,Mooij2,Fazio,WeesDual}], dual
to the BKT in the superconducting state. In the high temperature
phase, $T>T_{\scriptscriptstyle SI}$ the charges of either sign form
a gas of free 2$e$ charge Cooper pairs. At low temperatures,
$T<T_{\scriptscriptstyle SI}$, the charges of the opposite signs are
bound in dipoles. This charge binding-unbinding BKT is a realization
of the earlier theoretical observation that in a two-dimensional
electrolite with the logarithmic interaction between ions, the
transition at which the ions of the opposite sign become bound in
pairs, occurs upon lowering down the temperature of the
system~\cite{BKT-Prager}.

Having established a background, we are now in a position to give a
qualitative picture of the transport in a large JJA. The electric
field induced by the charge placed on a superconducting island (or
distributed over several islands) remains trapped within the JJA.
Thus a JJA is a one- or two-dimensional system with respect to the
electric field distribution.  Let us consider the situation where
the screening length, $\lambda_c$, appearing due to capacitance to
the ground exceeds the sample size $L$. In this case the energy
necessary to place an additional Cooper pair into the system is
$\Delta_c=E_c(L/d)$ in one- and $\Delta_c=E_c\ln(L/d)$ in
two-dimensional case.  This Coulomb energy can be presented as
$\Delta_c=(2e)^2/2C_{tot}$, where $C_{tot}$ is the total capacitance
of the array.  In an one-dimensional regular array $C_{tot}=C/N$
(the total capacitance for the system comprised of $N$ capacitors in
series).  Correspondingly, in a two-dimensional system
$C_{tot}\simeq C/\ln N$ for the array containing $N\times N$
junctions.  The thermally activated transport is thus governed by
the activation barrier $\Delta_c$ and the corresponding resistance
is

\begin{equation}
R\propto\exp(\Delta_c/T)\, , \label{NormIns}
\end{equation}
reproducing the experimentally observed dependence (1) in the Cooper
pair insulating state. The thermally activated resistance exists
only in a moderate temperatures region $\Delta_c>k_BT>E_c$, above
the transition temperature $T_{\scriptscriptstyle SI}$, 
where the free charges can
propagate across the array. To understand the dynamics below the
$T_{\scriptscriptstyle SI}$, let us notice first that the thermally
activated behavior (\ref{NormIns}) means that the whole array acts
in a synchronized manner as a one single superconducting island with
the characteristic capacitance $C_{tot}$.  In the insulating state
$E_c\gg E_{\scriptscriptstyle J}$ the charges at every junctions are
fixed and thus by the quantum mechanical uncertainty principle the
corresponding phases fluctuate loosely and so do the related local
electric fields.  However, as the current starts passing the array, 
all the internal phases synchronize in order to minimize the Joule losses.  
Thus the phase evolves coherently over the array implying that the whole system
behaves as a single superconducting island.

As a next step, one can realize that the Cooper pair propagation
across the array can be viewed as a propagation of a charge
soliton which is not necessarily confined to a one island.
Following\,\cite{VinNature} we introduce the local charge
density, $n_s(\textbf{r})$, which is normalized to give the total
soliton energy as $\Delta_c = E_c \int
d\textbf{r}n_s^2(\textbf{r})$. The probability for such a local
density to appear at point $\textbf{r}$ is proportional to
$\exp[-n_s^2(\textbf{r})/(2\langle\delta n^2\rangle )]$, where 
$\langle\delta n^2\rangle$ is
the mean square fluctuation of the local charge density. The
Cooper pair current is proportional then to the number of solitons
generated per unit time and traversing the array. The latter is
proportional to a product of all the above local probabilities at
all the points of the system: $\aleph_s\propto
\prod_\textbf{r}\exp[-n_s^2(\textbf{r})/(2\langle\delta n^2\rangle)]
=\exp\{-[1/(2\langle\delta n^2\rangle)]\int
d\textbf{r}n_s^2(\textbf{r})\} = \exp[-\Delta_c /(2E_c\langle\delta
n^2\rangle)]$. At temperatures above the charge binding-unbinding
transition, $T_{\scriptscriptstyle SI}\simeq E_c /k_B$, the
solitons are unbound and, according to the equipartition theorem,
$\langle\delta n^2\rangle = k_BT / E_c$, giving rise to thermally
activated resistance $R\propto\exp(\Delta_c /(k_BT))$. At low
temperatures, $T < T_{\scriptscriptstyle SI}$, the charge solitons
and antisolitons are bound, and therefore $\langle\delta
n^2\rangle$ is the probability of breaking these pairs, i.e.
$\exp(-E_c /(k_BT))$. This yields a double-exponential resistivity
in the superinsulating phase:
 \begin{equation}
   R\propto\exp\left[\frac{\Delta_c}{E_c}\exp\left(\frac{E_c}{k_BT}\right)
   \right] . \label{superinsulator}
 \end{equation}
The transition from the thermally activated insulating to superinsulating
behavior can be viewed as a manifestation of the fact that
$\langle\delta n^2\rangle$ represents the mean filling density
$\overline{n}$ for the energy state $E=E_c$ by the Cooper pairs. The
filling density $\overline{n}$, in its turn, is given by the Bose
statistics: $\delta n^2\equiv\overline{n}=[\exp(E_c/k_BT)-1]^{-1}$.

The outlined picture of the Cooper pair transport implies that the
internal part of an array acts coherently as a single
superconducting island, while the most of the applied voltage drops
at the leftmost and rightmost junctions. In other words the system
can be viewed as a two-junction system with the capacitance between
the central island and leads equal to the total capacitance of the
array.  The criterion for this scenario to hold at temperatures
$T>E_c/k_B$ can be presented as [see Eq.~(\ref{Condition}) above]:
      \begin{equation}
         N\left(\frac{E_{\scriptscriptstyle
         J}}{E_c}\right)^2\exp\left(-\frac{\Delta_c}{k_BT}\right)\ll 1\,
         .~\label{Condition1}
      \end{equation}
One sees that in a one-dimensional system, where $\Delta_c\simeq
NE_c$, the larger the system, the better the criterion
(\ref{Condition1}) is satisfied in compliance with the experimental
observation of~\cite{Haviland1} that the insulating behavior of the
1D Josephson array becomes more pronounced with the increase of the
system length.  In the 2D case the situation is more complicated,
but this criterion is satisfied pretty well at low enough
temperatures $k_BT\gtrsim E_c$.  In the superinsulating phase,
$k_BT< E_c$, the corresponding criterion following from
Eq.(\ref{Condition}) is met very well.

We expect that the model of the large Josephson junction array
applies fairly well to thin films in the critical region of the
superconductor-to-insulator transition, since as we have already
noticed, the internal structure is irrelevant with respect to
Coulomb properties of the system.

In conclusion, we have shown that the Cooper pair transport in the
insulating state of one- and two-dimensional Josephson junction
arrays is governed by the macroscopic Coulomb blockade.  The
macroscopic Coulomb blockade energy $\Delta_c\simeq E_c(L/d)$ in 1D
and $\Delta_c\simeq E_c\ln(L/d)$ in 2D systems. We have shown that
the charge binding-unbinding BKT-like transition separates the
\textit{insulating} high temperature state with the thermally
activated conductivity from the low temperature
\textit{superinsulating} state.  We have determined the conditions
under which the macroscopic Coulomb blockade is realized and the
conditions under which the Coulomb blockade activation energy
exhibits the system size dependence.

The questions that remain open include:
\begin{enumerate}
\item{The contribution of the quasiparticle current into the transport 
properties of large Josephson junction arrays.}
\item{The role of quantum fluctuations in  
the transport properties of large Josephson junction arrays.}
\end{enumerate}
These and related topics will be a subject of forthcoming
publication.
\section*{Acknowledgements}
We are deeply indebted to Yu. Galperin for most useful discussions.
This work was supported by the U.S. Department of Energy Office of
Science through contract No. DE-AC02-06CH11357, SFB 491 and
Alexander von Humboldt Foundation (Germany), and RFBR Grant
No.\,06-02-16704.  The authors will be most grateful for sending
copies (in the pdf format) of all the relevant works
that were overlooked in this paper, so that we could properly refer
to them in our forthcoming publications\cite{mail}.

\bigskip

\section*{APPENDIX}

As an illustration, we calculate the correlation function for a
single JJ case with the Hamiltonian

$$H_0=\frac{\hbar^2}{4E_c}\dot\varphi^2\,,$$
where $\varphi$ is the Josephson phase. In this case,
$$
K(\tau)=\int D\varphi
e^{i[\varphi(\tau)-\varphi(0)]}\exp\bigg[-\frac{\hbar}{4E_c}\int_0^{\hbar/(k_BT)}d\tau
\dot{\varphi}^2\bigg]~,
$$
The calculation is done via expanding $\varphi(\tau)$ into a Fourier
series:
$$
\varphi(\tau)=\sum_{\omega_n=2\pi k_B Tn/\hbar}e^{i\omega_n
\tau}\varphi_n\,.
$$
Replacing the functional integration over $\varphi(\tau)$ by
integration over Fourier coefficients $\varphi_n$ yields
\begin{eqnarray}\nonumber
K(\tau)=\prod_n \int d\varphi_n
\exp\bigg\{-\sum_n\bigg[\frac{\hbar^2 \omega_n^2 \varphi_n^2}{4E_c
k_BT}+\varphi_n(e^{i\omega_n \tau}-1)\bigg]\bigg\}\\
=\exp\big[-\sum_n \frac{4E_c k_B T}{\hbar^2
\omega_n^2}\sin^2(\omega_n\tau/2)\big]\nonumber\,.
\end{eqnarray}
In the interval $0<\tau<\hbar/(k_BT)$ the sum over $n$ yields:
$$
\sum_{n}\frac{4E_ck_BT\sin^2(\omega_n\tau/2)}{\hbar^2\omega_n^2}=
\frac{E_c}{\hbar}\tau-\frac{E_ck_BT}{\hbar^2}\tau^2,
$$
and the correlation function assumes the form
$$
K(\tau)=\exp\bigg(-\frac{E_c\tau}{\hbar}+\frac{E_ck_BT\tau^2}{\hbar^2}\bigg)~.
$$
Finally, the changing $\tau$ to $it$ gives
$$
K(t)=\exp\bigg(-i\frac{E_ct}{\hbar}-\frac{E_ck_BTt^2}{\hbar^2}\bigg)~.
$$
This result coincides with the one obtained by direct calculations
using the quantum-mechanical definition of $K(t)$ (Ref.\,[\onlinecite{FickSau}]) and
(\ref{corrfunction}).


\begin{thebibliography}{99}
\bibitem{comment4} Notice here, that the method of the analytic
continuation used in order to obtain the Eq. (\ref{korrfunctlargeT})
leads to the identical results as the direct change $\tau$ to $it$
in the Eq. (\ref{korrfunctlarge-imagT}). Moreover, the same method
of the analitic continuation, determined by (\ref{Fouriertransform})
and (\ref{Fouriertransform-new}), allows to obtain the Eq.
(\ref{Continuation}) in the general case as the nonzero winding
numbers are taken into account.
\end{thebibliography}
\end{document}